\documentclass[pre,8.5pt]{article}

\usepackage[super,sort&compress,comma]{natbib} 

\usepackage{sectsty}
\usepackage{balance} 
\usepackage{verbatim} 
\usepackage{a4wide}
\usepackage{amsmath} 
\usepackage{graphicx} 
\usepackage{lastpage} 
\usepackage{color}

\begin{document}

\newcommand{\T}{\mathrm{T}}
\newcommand{\dg}{\,^\circ}

\noindent\LARGE{\textbf{Spatially heterogeneous dynamics in a thermosensitive soft suspension before and after the glass transition}}
\vspace{0.6cm}

\noindent\large{\textbf{R\'emy Colin\textit{$^{a}$}, Ahmed M. Alsayed\textit{$^{b}$}, Jean-Christophe Castaing\textit{$^{b}$}, Rajesh Goyal\textit{$^{b}$}, Larry Hough\textit{$^{b}$} and
B\'ereng\`ere Abou\textit{$^{a}$}}}\vspace{0.5cm}

\noindent \normalsize{The microscopic dynamics and aging of a soft
  thermosensitive suspension was investigated by looking at the
  thermal fluctuations of tracers in the suspension. Below and above the
  glass transition, the dense microgel particles suspension
  was found to develop an heterogeneous dynamics, featured by a
  non Gaussian Probability Distribution Function (PDF) of the probes' displacements, with an exponential tail. We show that non Gaussian shapes are a characteristic of the ensemble-averaged PDF, while local PDF remain Gaussian. This shows that the scenario behind the non Gaussian van Hove functions
  is a spatially heterogeneous dynamics, characterized by a
  spatial distribution of locally homogeneous dynamical environments through the sample, on the considered time scales. We characterize these statistical
  distributions of dynamical environments, in the liquid, supercooled, and glass states, and show that it can
explain the observed exponential tail of the van Hove functions observed in the concentrated states. The intensity of spatial heterogeneities was found to amplify with
increasing volume fraction. In the aging regime, it tends to increase as the glass gets more
arrested.
 }
\vspace{0.5cm}

\section{Introduction}

\footnotetext{\textit{$^{a}$~Laboratoire Mati\`ere et Syst\`emes Complexes, UMR
    CNRS 7057 \& Universit\'e Paris Diderot, 10 rue A. Domon et L. Duquet, 75205 Paris Cedex 13, France. E-mail: remy.colin@univ-paris-diderot.fr; berengere.abou@univ-paris-diderot.fr;}}
\footnotetext{\textit{$^{b}$~Complex Assemblies of Soft Matter Laboratory, UMI CNRS 3254, Rhodia INC., 350 G. Patterson Blvd, Bristol PA 19007, USA. }}

Dense colloidal suspensions are a powerful model system to study the
glass and jamming transitions. With increasing volume fraction, the
suspension dynamics may experience a slowing down of several orders of
magnitude, as well as a change of qualitative behaviour, from viscous
to elastic, at reasonable experimental time scales, with no
significant structural changes in the
suspension\cite{nature-Pusey-1986,Science-vanBlaaderen-1995}. This
colloidal glass transition is analogous to the glass transition in
liquids and polymer melts and to the jamming transition in granular
materials\cite{Nat-Liu-1998}. Because the colloidal particles are
larger than molecules, these systems provide slower accessible times
scales than in atomic and molecular glasses, and the possibility to be
probed with optical techniques including microscopy and dynamic light
scattering\cite{Cipelletti-Weeks}. Hard spheres suspensions have been extensively investigated for their
relative simplicity in modelling the glass transition\cite{nature-Pusey-1986}. Recently, soft
colloidal suspensions have also attracted attention. They can be
flexibly designed with various particles softness and interactions --
attractive or repulsive -- allowing for a wider variety of behaviour in
mimicking the glass transition in traditional glasses\cite{Mattson-Nature-2009}. One of
the most investigated soft systems consists of aqueous suspensions of
thermosensitive microgels, made with amphiphilic cross-linked
poly(N-isopropylacrylamid) polymer (pNIPAm)\cite{ACIS-Saunders-1999}. These particles can be
designed with various softnesses, depending on the cross-linkers
density\cite{CPS-Senff-2000}. Moreover and not least, the particles
radius in the suspension, and therefore their volume fraction, can be reversibly tuned with temperature, which provides a unique way
to explore the phase behaviour on the same sample. This flexibility
has been used to study the phase and behavioural
transitions of colloidal suspensions\cite{Langmuir-Stieger-2004}, such
as crystal formation\cite{CPS-Hellweg-2000,JPCB-Lyon-2004} or
melting\cite{Science-Alsayed-2005,PRE-Han-2008}, organisation in
constrained geometry\cite{PRE-Lohr-2010}, and glass and jamming
transitions\cite{PRL-Purnomo-2008,Nature-Zhang-2009,PRE-Purmono-2010}.

Colloidal glasses are intrinsically out-of-equilibrium systems which
age~: their physical properties depend on the time elapsed since they
were formed. When approaching the glass transition, colloidal
suspensions exhibit heterogeneous
dynamics\cite{Langmuir-Kasper-1998,PRE-Marcus-1999,Science-Kegel-2000,Science-Weeks-2000,PRL-Cloitre-2000,PRE-Lynch-2008,PRL-Yunker-2009},
also observed in other glass forming
systems\cite{PRB-Zorn-1997,AnnRevGlass-Ediger-2000,Nature-Vidal-2000,PRL-Marty-2005,PRE-Abate-2007}
and in simulations\cite{PRL-Doliwa-1998,
  JNCS-Glotzer-2000,PRE-Elmasri-2010}. It is characterized by non
Gaussian tails of the van Hove correlation functions, and is now
admitted to be a characteristics of the glass and jamming
transitions\cite{JPCM-Richert-2002,PRL-Chaudhuri-2007}. On a
theoretical point of view, many interpretations were proposed to
capture the heterogeneous
dynamics\cite{JNCS-Richert-1994,PRE-Vorselaars-2007,AnnRevGlass-Ediger-2000}. Experimentally,
a connection with cage motion rearrangements with string like motions
of high mobility, was established by different
groups\cite{PRE-Marcus-1999,Science-Kegel-2000,Science-Weeks-2000}. Recently,
it was related to a double Gaussian in a soft thermosensitive glass,
suggesting two distinct dynamic populations of
particles\cite{PRE-Purmono-2010}. In another recent investigation in
thermosensitive suspensions, multiple regimes of dynamic
heterogeneities were found with increasing volume fraction
\cite{Sessoms-2009}. While widely investigated, the scenario of
dynamic heterogeneity in glass forming materials, still remains
elusive.

In this paper, we report an experimental investigation of the
heterogeneous dynamics in a thermosensitive soft spheres suspension
when approaching the glass transition from below (increasing volume
fraction), and beyond. Our system consists of a suspension of pNIPAm
microgel particles, with soft repulsive potential interactions, with
temperature as the external control parameter of the volume
fraction. We first investigate the aging dynamics of the soft glass by
looking at the thermal fluctuations of the tracers. We then
characterize the dynamic heterogeneity in the suspension by computing
the ensemble-averaged and the local Probability Distribution Functions
(PDF) of the probes' displacements. We characterize the statistical
distributions of the dynamical environments explored by the
tracers. At equilibrium in the supercooled regime, and in the glass
state, we show that it can explain the exponential tail of the van
Hove functions. Finally, the intensity of the spatially heterogeneous
dynamics was investigated with increasing volume fraction, below and
above the glass transition.

\section{Material and Methods}

\subsection{Soft microgel particle suspensions}

Our system consists of a suspension of micron size cross-linked pNIPAm
microgel particles, synthesized as follow : $20$~mg of the
cross-linker BIS (N,N'-Methylenebisacrylamide, Polysciences, Inc.),
$20$~g of NIPA (N-isopropylacrylamide, Polysciences, Inc.), and
$400$~ml of deionized water are loaded into a special three-neck flask
equipped with a stirrer, thermometer, and a gas inlet. The resultant
mixture is stirred, heated at $82 \dg$C, and bubbled with dry nitrogen
for $10$ min to remove dissolved oxygen. A solution of $200$~mg of APS
(Ammonium Persulfate) dissolved in $1$~ml of deionized water is then
added to the mixture to start the polymerization reaction. The mixture
is continuously stirred at $82 \dg$C for $30$ min and then allowed to
cool down to room temperature. The resultant particles are centrifuged
and resuspended in deionized water few times to remove unreacted
monomer, homopolymers, and other salts.

The microgel particles are swollen at low temperatures, $\mathrm{T} <
30\dg$C, and collapsed above $\T > 35 \dg$C, as shown in Figure
\ref{DT}. Below $\T=32 \dg$C, where we investigate the system
behaviour, they have soft repulsive interactions. The hydrodynamic
radius of the particles is a monotonously decreasing function of
temperature, the LCST being around $\T = 32\dg$C. In colloidal hard
spheres suspensions, the phase behaviour is controlled by the volume
fraction. Concentrated colloidal hard spheres suspensions are known to
enter an out-of-equilibrium glass state above a glass transition
volume fraction around $\Phi_g \sim
0.58$\cite{nature-Pusey-1986}. Here, the volume fraction can be
monitored with the temperature, allowing for an exploration of the
phase diagram on the same sample. The polydispersity, of the order of
$10\%$ for our samples, as well as the particles softness, suppresses
the crystallisation at high volume fractions (obtained at low bath
temperatures). Starting from a liquid state and increasing volume
fraction, the suspension was found to enter a glass state,
characterized by aging over long time scales.

A concentrated stock suspension of pNIPAm particles was
prepared. Polystyrene beads with a diameter of $1\, \mu$m were
dispersed in the pNIPAm suspension to a volume fraction of
$0.08\%$. The non-index-matched beads were used as probes of the
thermal fluctuations in our samples. The suspension was then injected
in a square chamber ($3 \times 3$ mm$^2$), made of a microscope plate
and a cover-slip separated by a thin adhesive spacer ($250\,\mu$m
thickness). The chamber was then sealed with glue in order to avoid
contamination and evaporation.

The microgel particles suspension at $\T=30 \dg$C, was found to behave
as an equilibrium 'liquid' state, characterized by the nearly linear
dependency of the probes mean-squared displacement with the lag time
(see section \ref{aging}). It was thus taken as our reference liquid
equilibrium state at volume fraction denoted $\Phi_0$.

An estimate of the lower bound for the sedimentation time of the Latex
probes over $1$ micrometer is $200$ s, in the reference liquid of
viscosity equal to $6$ mPa.s (taking the water density as the
suspension density). The sedimentation is thus slow enough to allow us
recording during few minutes. Generally, the sample is left on the
microscope at $\T=30 \dg$ C during two minutes for recording and then
quenched at low temperature to reach high volume fraction. At high
volume fraction, in the glass state, the sedimentation of the probe
particles is negligible due to the high viscosity of the material, and
recording for hours is possible.

In soft spheres suspensions, the volume of the microgel particles
might not be fixed at different volume fractions, especially at high
volume fractions, and the particle number density is no longer
equivalent to the volume fraction, the control parameter of colloidal
suspensions. Moreover, the shape of the deformable microgel particles
is not fixed, and high volume fraction suspensions can still be fluid
above random close packing since particle deformations allow for
rearrangements\cite{Mattson-Nature-2009}. Here determining the volume
fraction $\Phi_0$ in the reference state, with the Einstein's
viscosity relation, gives us a value larger than unity of poor
relevance, that indicates an already compressed state, and extra shrinkage of the particles. Standard
methods, such as counting (nearly index-matched particles with the
solvent) or weighting, were not possible. In the following, because we
are interested in the phase behaviour with increasing or decreasing
the control parameter, the volume fraction of the suspensions at
equilibrium at $\T=30 \dg$ C will be expressed in terms of $\Phi_0$ at $\T=30 \dg$ C. Increasing
volume fraction will be achieved by quenching the sample from the
reference state at $\Phi_0$ and $\T=30 \dg$ C, resulting in a new couple ($\Phi$, T) and the corresponding phase
behaviour will be explicited.

\subsection{Dynamic Light Scattering}
\label{DLS}	

Dynamic Light Scattering (DLS) experiments were performed on low
concentration suspensions of microgel particles.  The set-up is
composed of a MELES GRIOT He-Ne laser source, operating at the
wavelength $\lambda = 632.8$ nm, a Brookhaven Instrument
Corp. automatic goniometer furnished with a BI-9000AT digital
autocorrelator.

In our slightly polydisperse suspensions, analyzing the time
dependence of scattered light was shown to be less susceptible to
experimental fluctuations than analyzing the angular dependence of the
scattered light \cite{Berne_Pecora}. The recorded autocorrelation
functions was analysed using a William-Watt fit of the first
scattering function~: \[g_1(t) = e^{-(t/\tau)^\beta}\] where $\tau$ is
the decay time and $\beta$ takes into account the slight
polydispersity of the sample. The exponent $\beta$ was found between
$0.8$ and $1$, corresponding to a polydispersity between $5\%$ and
$15\%$.

\begin{figure}
\center
\includegraphics[width=9.0cm]{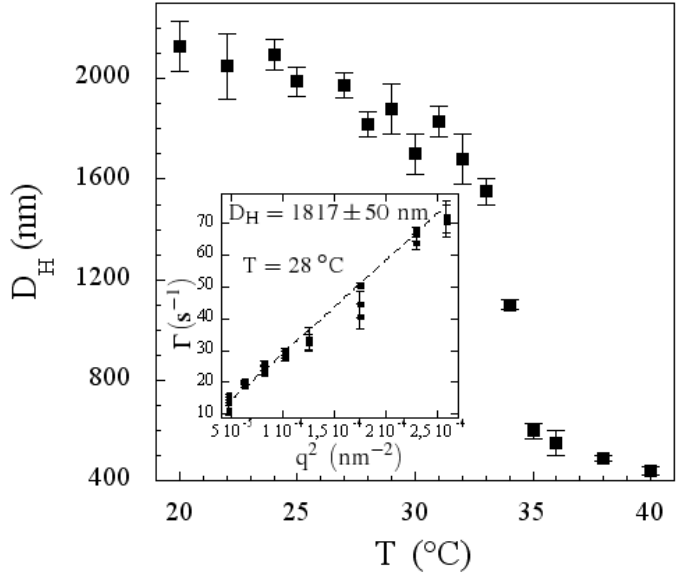}
\caption{Mean hydrodynamic diameter of the pNIPAm particles as a
  function of temperature, measured with dynamic light scattering
  experiments. The polydispersity was found to vary between $5\%$ to
  $15\%$, depending on the realization. A sharp decrease of the
  microgel particles size was observed around
  $\T=32\dg$C. Their diameter was multiplied by $5$ from
  the fully collapsed state at high temperature to the fully swollen
  state at low temperature, reflecting the small degree of
  cross-linking. Inset: linear relation between the inverse $\Gamma$
  of the correlation function decay time $\tau$ and $q^2$, here at
  $\T=28\dg$C. The linear dependency validates that only the diffusion of the particles is probed, and allows us to determine
  the particles diameter.
}
\label{DT}
\end{figure}

Here, DLS experiments were also performed
at different scattering angles $q$ for even more dependable
results. The decay time $\tau$ was
shown to scale with the scattering wavenumber $q$ according to
$\frac{1}{\tau} \propto q^2$, indicating a diffusive behaviour of the
scatterers (Fig. \ref{DT}, inset). For each temperature, the mean hydrodynamic diameter of the particles $D_H$ could then be
deduced from the relation : \[\frac{1}{\tau} = \frac{k_B
  \mathrm{T}}{3\pi \eta(T) D_H} q^2\] where $k_B$ is the Boltzmann
constant, $T$ the temperature, and $\eta(T)$ the solvent viscosity.
Figure \ref{DT} shows the monotoneous decrease of the mean
hydrodynamic diameter as a function of temperature. A sharp decrease
of the particle size occurs around $T= 32\dg$C.

In these DLS experiments, we are aware that we are not in the
conventional range of radii $R$ for using DLS for particle sizing,
since Mie scattering theory applies ($q R\sim 1$), and internal
fluctuations of the particles might be probed. However, the good
agreement of the $1/\tau \sim q^2$ law indicates a diffusive behaviour
of the microgel particles, giving evidence that the internal
fluctuations do not contribute to the scattered intensity. We also checked that we could obtain a good estimate of the hydrodynamic
diameter for calibrated polystyrene microbeads of $1\,\mu$m and
$1.8\,\mu$m in diameter, within a $5\%$ error, in the same range of
concentration. We are thus
confident in our results, within an error bar of $5$ to $10\%$.

\subsection{Particle Tracking}
	
The thermal motion of the tracers was recorded using a inverted Leica
DM IRB microscope with a $\times 100$ oil immersion objective, coupled
to a CMOS camera (Phantom v9). The camera was typically running at
$20$ frames per second (fps) in the more concentrated states, and at $500$ fps  in the low viscosity suspensions, with a $56\times 56$ $\mu\text{m}^2$
field of view. Most of the records were $2\,\cdot 10^3$ frames
long. Depending on the realization, $10$ to $70$ probe thermal motions
could be recorded in the same field of view. For reliable
analysis of the Brownian motion, particular attention was brought to
record the motion far from the rigid wall imposed by the glass
slides. 

A home-made image analysis software allowed us to track the bead
positions $x(t)$ and $y(t)$ close to the focus plane of the
objective. For each probe $i$, the time-averaged
mean-squared displacement (MSD), $\langle\Delta r^2_i (t)
\rangle_{t'}= \langle[x_i(t' + t) - x_i(t' )]^2 + [y_i(t' + t) - y_i(t'
  )]^2 \rangle_{t'} $ was
calculated, improving the statistical accuracy. When necessary, an ensemble-averaged MSD over
all the probes, denoted as $\langle\Delta r^2 (t) \rangle_{t',i}$, could
be computed. In the following, $\langle\Delta r^2_i (t)\rangle_{t'}$
will refer to the temporal-averaged MSD for one probe, and $\langle\Delta
r^2 (t)\rangle_{t',i}$ to the ensemble and temporal averaged MSD.

In our aging experiments, the recording time was short compared to the
aging dynamics, which means that no significative aging was observed
during the recording. The system could then be considered to be in a
quasi stationary state, at a given aging time, during the
recording. 
The resolution limit
of our set-up was tested using a highly concentrated ($95 \%$ wt)
glycerol solution, and was found to be $15$ nm.

Probability Distribution Functions (PDF) of the tracer
displacements (self part of the van Hove correlation function) were
also computed for each tracer : $p_i\left(\Delta x, t\right) = \langle
\delta \left(\Delta x - \Delta x_i\left(t\right)\right)\rangle_{t'}$,
and for the overall sample : $ p\left(\Delta x, t\right) =
\langle\delta \left(\Delta x - \Delta
x_i\left(t\right)\right)\rangle_{t',i}$, for different lag times $t$,
where $\Delta x$ denotes the elementary one-dimensional displacement
along the $x$ axis of the field of view and $\delta$ refers to the
delta function. In this study, the system is isotropic, and the
behaviour was found to be quantitatively the same on both $x$ and $y$
axis.

\subsection{Quenching procedure}
	
One important consideration for the investigation of the aging
dynamics in glasses is the preparation and reproducibility of the
initial glassy state. In colloidal glasses, one has to quench the
system from the liquid state by increasing rapidly the volume
fraction. In pNIPAM microgel particles suspensions, such a possibility
is offered by varying the temperature. This can be achieved rapidly
because the swelling / deswelling rate is of the order of the
millisecond \cite{JCP-Tanaka-1979}. After a quench, microgel particles which are
swollen in a few seconds, generate stresses through the sample, that
relax with time. 

The microscope objective temperature could be set between $ 20\dg
\rm{C}< T< 35\dg\rm{C}$, using a Bioptechs objective heater
coupled with a cooling device. The sample, kept at high temperature on
a Peltier plate at $\T=30\dg$C, was suddenly placed on the
objective, set at a lower temperature. The sample
temperature was then controlled through the oil immersion objective in contact. The
equilibration of the temperature sample with the immersion oil by
thermal diffusion was achieved in a few seconds\footnote{We checked that no
heat convection could take place in our samples. In this
  geometry, the Rayleigh number for water, the lower limit viscosity
  we used, is $1$ for a reasonable temperature difference of
  $10\dg$C, which is three orders of magnitude below the
  threshold of thermal convection.}.

All the quenches described here were performed from our equilibrium 'liquid'
state at $\Phi=\Phi_0$, $\T=30\dg$C, resulting in a new couple describing the sample ($\Phi$, T). The quench defines the origin of aging times $t_w$. At
$t_w=0$, the volume fraction of the suspension increases in a few
milliseconds, and for sufficiently high volume
fractions reached after the quench, a glass was formed out of the
initial equilibrium liquid. Aging, due to structural relaxations in the deformable soft spheres suspension, was observed.

The relative size of pNIPAm particles to probe particles changes by 20
$\%$, through the quench. As a result, it is not the same in the
starting liquid state and in the corresponding glass state. Moreover,
the temperature change slightly affects the particles softness. As for
the polydispersity, it does not change much between $\T=20 \dg$ C and
$\T= 30 \dg$C. As long as these quantities (relative size of pNIPAm
particles to probe particles, particles softness, polydispersity)
remain constant during the aging process, after the quench, it will
not affect quantitatively our results. It would be an issue to
determine quantitatively the phase diagram of such a suspension.

\section{Results and discussion}

\subsection{Aging dynamics in the glass}
\label{aging}

\begin{figure*}
\includegraphics[width=14cm]{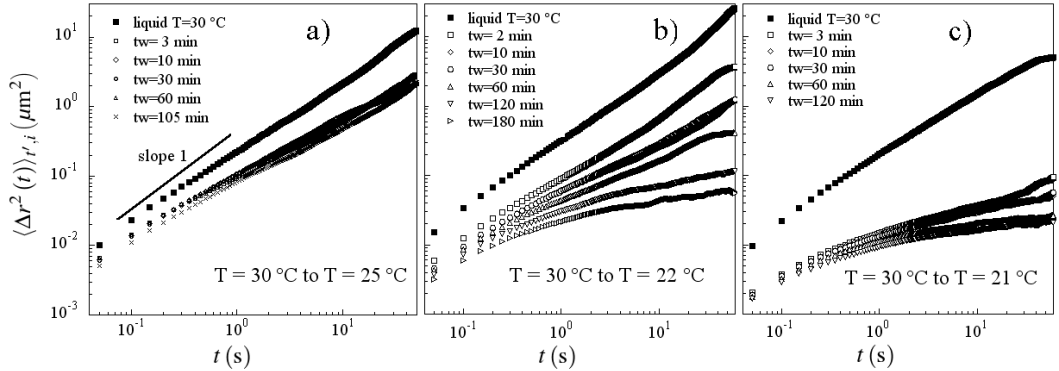} \centering

\caption{Mean-squared displacement of the tracers ($1\,\mu$m
  in diameter) as a function of the lag time in the pNIPAM microgel
  particles suspension, for quenches of different amplitude. Starting
  from an equilibrium liquid state at $\Phi=\Phi_0$, $\T=30\dg$C, the sample is
  quenched down to $\T=25,22$ and $21\dg$C, from the left to the
  right. \textbf{(a)}: at $\T=25\dg$C, the suspension exhibits an
  equilibrium visco-elastic behaviour. No aging was observed over more
  than one hour. The straight line represents the slope 1
  corresponding to a purely diffusive behaviour. \textbf{(b)}: at
  $\T=22\dg$C, the sample enters a glassy state and slowly ages
  over several hours. \textbf{(c)}: for a deeper quench at
  $\T=21\dg$C, the aging dynamics was found to be
  faster. \textbf{(b,c)}: upon aging, the MSD decreases continuously,
  and develops a relatively smooth 'plateau' for large lag times $t$,
  while it is still diffusive for short lag times. Moreover, the
  deeper the quench in the glassy state, the faster the glass aging
  dynamics, {\it i.e.} the faster and the deeper the plateau
  develops.}
\label{MSDN16}
\end{figure*}

The aging dynamics of the microgel particles suspension after a
quench in temperature, was investigated by looking at the
Brownian motion of the tracers. The mean-squared displacement (MSD) of
the probes as a function of the lag time, is displayed in
Fig. \ref{MSDN16}, in the reference 'liquid' state at $\Phi=\Phi_0$, $\T=30\dg$C, and after
quenches of different amplitudes.

In the reference equilibrium state, the MSD of the probes is nearly diffusive, characterized by a nearly linear dependency with the lag time, and no
aging could be observed over more than $4$ hours. All the quenches were performed from this 'liquid' state at
$\Phi=\Phi_0$, $\T= 30\dg$C. The suspension viscosity was estimated with the Stokes relation $\eta= 4 k_B T t/ (6\pi  R \langle\Delta r^2\left(t\right)\rangle_{t',i})= 6.1  \pm 1.1 \, \textrm{mPa}\cdot$s, where $R$ is the tracer radius and $k_B$ the Boltzmann constant.

For a quench from $\T= 30\dg$C to $\T=
25\dg$C, the suspension,  still at
equilibrium, becomes slightly visco-elastic, characterized by a non linear increase of the MSD with the lag time; no aging was observed over more than one hour
(Fig. \ref{MSDN16}(a)). For larger amplitudes quenches -- down to $\T=
22\dg$C and $\T= 21\dg$C -- the suspension was found to fall
out-of-equilibrium and to age over hours, indicating that the sample
has entered a glassy state, as shown in Fig. \ref{MSDN16} (b,c). Upon
aging, the MSD decreases continuously, and develops a subdiffusive
'plateau' at large lag times $t$, while it is still diffusive for
short lag times. Moreover, with
increasing the quench amplitude, the aging dynamics is faster and the
suspension goes in a deeper glass state. 

The underlying corresponding picture for the MSD data in the glass
state, is the following : the first diffusive regime, at short times,
is associated with the motion of the tracer in the cage formed by the
surrounding microgel particles. The second regime -- smooth plateau
becoming deeper upon aging -- reflects the elastic response of the cage at the
probe length scale. Upon aging, the cage becomes stiffer due to
structural relaxations of the soft deformable spheres, and the tracer
gets more trapped. Since
the pNIPAm particles are soft, the probe can penetrate the shell of
the surrounding microgel particles, and the plateau is a 'smooth' one. In these experiments, the cross over to a
diffusive motion of the tracers at larger lag times (cage
rearrangement) was not observed, indicating that the tracers remain
trapped on the observation time scale.

Here, the glass transition was defined
with respect to the out-of-equilibrium regime experienced by the
suspension above a certain threshold volume fraction. This is reasonable
since no crystal is expected because of the suspension polydispersity. The onset temperature below which the
suspension enters a glassy out-of-equilibrium state, was found between
$\T= 24 \dg$C and $\T= 22 \dg$C (corresponding to $ 1.5 \, \Phi_0<\Phi< 1.6\,  \Phi_0$). Given the relatively small dependence
of the microgel particles size as a function of temperature in this range (see Fig. \ref{DT}), the
'glass transition' temperature could not be determined with more
accuracy.  

\begin{figure}
\centering
\includegraphics[width = 8.0 cm]{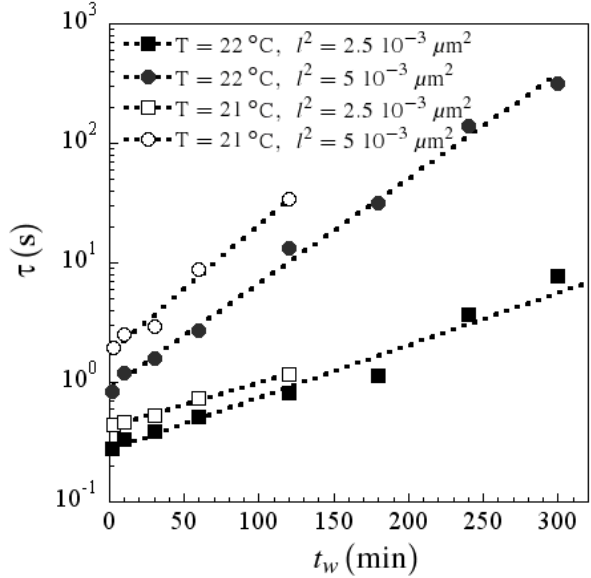}
\caption{Characteristic times $\tau$ versus aging time $t_w$ after quenches
  at $\T=22\dg$C and $\T=21\dg$C , recorded over 5 hours. Here, the characteristic time was defined as the lag time needed to
  reach a certain value $l^2$ of the MSD. The characteristic time was found to increase
    exponentially with aging time, such as $\tau \sim
\exp(k t_w)$.}
\label{tau}
\end{figure}

From the MSD data, a characteristic time $\tau$ associated with the
probes diffusion was extracted. This characteristic time was
defined as the lag time necessary for a probe to diffuse over a
lengthscale $l$, such as $\langle\Delta
r^2\left(t\right)\rangle_{t',i} = l^2$. Fig. \ref{tau} shows the
evolution of the characteristic time $\tau$ with aging time, for two
different values of $l^2$ -- $2.5\,10^{-3}\mu$m$^2$ and $5.0\,10^{-3}\mu\mathrm{m}^2$ --
after a quench from $\T = 30\dg$C to $\T = 22\dg$C. On
these length scales, typically of the order of one tenth of the probe
and pNIPAm particle radius, the characteristic time grows
exponentially with aging time by $2$ decades as $\tau \sim \exp(k t_w)$. Another definition involving
$Q\left(l,\tau\right) = \langle e^{-\Delta r_i^2 (\tau) /
  l^2}\rangle_{i,t'} = 0.6$, where $Q\left(l,\tau\right)$ is a
correlation function equivalent to the intermediate scattering
function, gave totally similar results \cite{epl-Lechenault-2008-1}. 

Such an exponential increase of the relaxation time was already
observed in other soft glasses and gels, often coupled to a transition
to an asymptotic aging regime ($\tau \sim t_w$)\cite{PRE-Abou-2001,
  PRE-Bellour-2003, PRE-Kaloun-2005}. Here the characteristic time we define is not a relaxation time in the usual definition. The
asymptotic aging regime might develop at later aging times $t_w$, on
the considered length scales. Experimentally, no cross over from the exponential growth to
the full aging regime could be observed, by changing the quench depth to
accelerate the aging process or measuring the characteristic time over
$7$ hours.

In these experiments, we probe the dynamical behaviour of the
microscopic environment at the scale of the pNIPAm particles, that
might be very different from larger scale response. The radius of the
probes being of the order of the radius of the pNIPAm particles, there
is no scale separation between the probes and the environment
constituting elements. We believe that the probes reflect the dynamics
of the surrounding pNIPAm colloidal particles, even though the
relation between the probes and microgels dynamics is not
straightforward and depends on a number of parameters, especially the
chemistry of the probes, and the probe size\cite{RPP-Waigh-2005}.

\subsection{Spatially heterogeneous dynamics}

\subsubsection{The spatial distribution of local Gaussian dynamics behind non Gaussian van Hove functions}
\label{sec-pdf}
\begin{figure*}
\centering
\includegraphics[width =14cm]{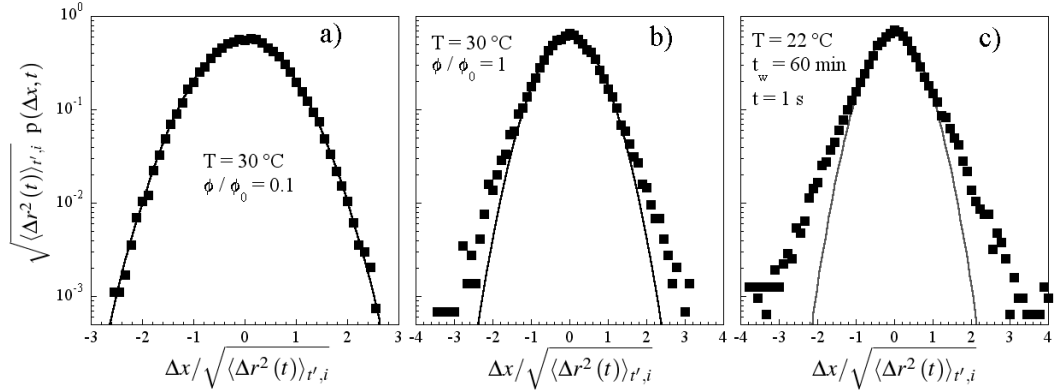}
\caption{Rescaled Probability Distribution Functions (PDF) of the
  probes displacement as a function of the rescaled
  displacement. \textbf{(a)} in the diluted suspension at equilibrium at $\Phi=\Phi_0$, $\T=30\dg$C,
  for a lag time $t=0.1$~s. In abscisse, the unity corresponds to
  $\Delta x = 0.6 \, R $, where $R$ is the probe radius; \textbf{(b)}
  in the suspension at $\T=30\dg$C, for a lag time $t=0.15$~s. In
  abscisse, the unity corresponds to $\Delta x = 0.38 \, R $;
  \textbf{(c)} in the glass state after a quench from $\T=30^\circ$C
  to $\T=22\dg$C, at aging time $t_w=60$ min, for a lag time
  $t=1$~s. In abscisse, the unity corresponds to $\Delta x = 0.42 \, R
  $. The black curves are Gaussian fits $Y = \exp\left(-X^2\right) /
  \sqrt{\pi}$. In the diluted suspension, the PDF is a Gaussian, as
  expected in a purely viscous liquid. In the suspension at
  $\T=30\dg$C, the PDF slightly departs from a Gaussian at large
  displacements. In the glass state, the PDF exhibits a central
  Gaussian part for short displacements, and an exponential like decay
  for large displacements, observed at any lag time and aging time. }
\label{probaN16}
\end{figure*}

We now analyze the probe thermal fluctuations by looking at the
Probability Distribution Functions (PDF) of the probes' displacements
(or van Hove functions). Near the glass transition, they have been
shown to exhibit non Gaussian shapes, usually assigned to 'dynamic
heterogeneity'. Here, we analyze the ensemble-averaged and individual PDF of the tracers.

Figure \ref{probaN16} shows rescaled ensemble-averaged PDF
$\sqrt{\langle\Delta r^2 \left(t\right) \rangle_{t',i}} \:\: p(\Delta
x,t)$ as a function of the normalized displacement $\Delta x /
\sqrt{\langle\Delta r^2 \left(t\right) \rangle_{t',i}}$ in the
equilibrium liquid state at different volume fractions, and in the
glass state. Experiments performed in the diluted
suspension at
$\Phi=0.1 \, \Phi_0$, $\T=30\dg$C (dilution factor $10$, purely viscous, viscosity $1.85 \pm 0.15$~mPa.s), shows a Gaussian PDF, as expected in a viscous
homogeneous fluid (Fig. \ref{probaN16}(a)). In the equilibrium reference state
at $\Phi=\Phi_0$, $\T=30\dg$C, the ensemble-averaged PDF were found to be {\em nearly}
Gaussian at any lag time (Fig. \ref{probaN16}(b)), while in the glassy
state, they are Gaussian for short displacements, and exhibit an
exponential like decay for large displacements, at any lag time
allowing for a good statistics ($t<2$~s), and aging time
(Fig. \ref{probaN16}(c)). As shown here, the shape of the
ensemble-averaged PDF changes with increasing volume fraction~: it
goes from a pure Gaussian at low volume fraction to a PDF with a
Gaussian central part for short displacements, and a broad exponential
like decay at large displacements, at higher volume fraction. Moreover, the tail of the distribution becomes more pronounced when
approaching the glass transition (increasing volume fraction).

\begin{figure}
\includegraphics[width = 8.7cm]{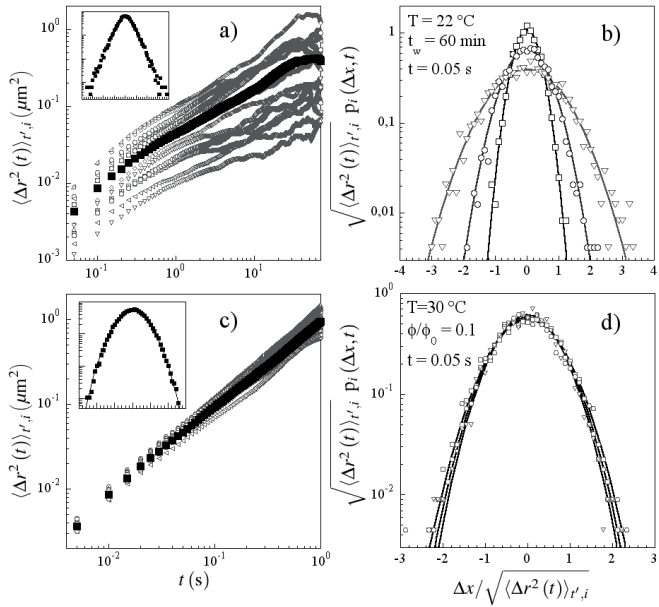}
\centering
\caption{\textbf{(a)}: MSD as a function of the lag time for different
  probes (open symbols) and the ensemble-averaged MSD over the probes
  (large filled symbols), in the glass state at aging time $t_w=60$
  min, after a quench from $\T=30\dg$C to $\T=22\dg$C. A broad
  spatial distribution of MSD was observed for each $t$. In inset, the
  corresponding ensemble-averaged PDF is plotted. \textbf{(b)}:
  Individual PDF of the displacements for three probes in the glass
  state after a quench from $\T=30\dg$C to $\T=22\dg$C, for a
  lag time $t=0.05$ s, at aging time $t_w=60$ min. For each probe, the
  individual PDF is a Gaussian with a different width, resulting from
  a different MSD. \textbf{(c)}: the narrow spatial distribution in
  the diluted liquid state at $\Phi=0.1 \,\Phi_0$, $\T=30\dg$C, with the corresponding
  Gaussian ensemble-averaged PDF in inset. \textbf{(d)}: the
  individual particles' displacements distributions are Gaussian, with
  the same widths. }
\label{probaN16expl}
\end{figure}

We now consider the individual motions of the probes through the
sample. In the glass state, where the ensemble-averaged PDF were found
to depart from a Gaussian, the spatial distribution of the MSD is wide
(Fig. \ref{probaN16expl}(a)). The MSD of the individual probes are
drastically different from one probe to another, even at shortest lag
times where the statistics is the highest. The corresponding
individual PDF of the probes' displacements are Gaussian, $p_i(\Delta
x, t) \sim \exp{(-(\Delta x(t))^2 / \Delta_i)}$, at any lag time
allowing for a good statistics ($t<2$~s) and aging times $t_w$. Their variance $\Delta_i$ were found to
be widely distributed through the glass sample (Fig. \ref{probaN16expl}(b)). The resulting
ensemble-averaged PDF over all the probes, in the top inset and
already shown in Fig. \ref{probaN16}(c), exhibits a central Gaussian
part, and an exponential like decay at large displacements. By
comparison, in the diluted equilibrium liquid state ($\Phi=0.1 \,
\Phi_0$, $\T=30\dg$C), the MSD are narrow distributed
(Fig. \ref{probaN16expl}(c)). The individual PDF of the probes are
Gaussian and similar in width (Fig. \ref{probaN16expl}(d)), and the
resulting ensemble-averaged PDF, shown in the bottom inset and in
Fig. \ref{probaN16}(a), is a Gaussian. In the reference equilibrium
state (not shown here, $\Phi= \Phi_0$, $\T=30\dg$C), the distribution
of the individual MSD through the sample is clearly narrower than in
the glass state, but wider than in the diluted liquid state. The
individual PDF are Gaussian with slightly different widths, and the
corresponding PDF slightly departs from a Gaussian at large
displacements, as shown in Fig. \ref{probaN16}(b). 

In the Appendix, we analyze the ensemble-averaged and individual
Cumulative Distribution Function (CDF) of the probes' displacements in
the liquid state and in the glass state, in order to clearly identify
the shape of the PDF at short and large displacements. Using the CDF
to graph the displacements distribution of sample data and try to
identify it with respect to known distributions avoids problems of bin
size and bin location. We could validate that the ensemble averaged
PDF is Gaussian for short displacements, and that its decay at large
displacements is well adjusted by an exponential over at least 3
decades in the glass state, while local PDF remain Gaussian (see Appendix).

In our system, the individual distributions of the probes'
displacements were found to be Gaussian, whether at equilibrium or
out-of-equilibrium, even though non diffusive, in the considered range
of lag times $0.05<t<2$~s, at any volume fraction or age. They are
given by $p_i(\Delta x,t)\sim \exp{(-(\Delta x(t))^2 / \Delta_i)}$, with
a variance $\Delta_i$ which becomes widely distributed through the
sample when approaching the glass transition. The variances $\Delta_i$
characterize the local dynamical environment\footnote{Note that the
  variance $\Delta_i$ of the individual PDF is by definition directly
  related to the individual MSD, by $\Delta_i(t)=\langle\Delta
  r^2_i\left(t\right) \rangle_{t'}$. }. They can also be considered as
a measure of the 'probe mobility' within the suspension, since the
larger the variance $\Delta_i$, the more 'mobile' the probe is. 

The scenario behind the non Gaussian van Hove functions is a spatially
heterogeneous dynamics of the suspension. At low volume fraction, the
spatial distribution of the probes' mobilities is narrow, revealing an
homogeneous suspension, and the ensemble-averaged PDF, $p(\Delta
x,t)$, is a Gaussian. With increasing volume fraction, the spatial
distribution of the mobilities gets wider, revealing a spatially
heterogenous dynamics. A sum of local Gaussians $p_i$ of different
widths is no longer a Gaussian. As a consequence, the corresponding
ensemble-averaged PDF deviates from a Gaussian. The observed non
Gaussian tails of the ensemble-averaged $p(\Delta x,t)$ are thus the
result of the wide spatial distribution of local dynamical
environments $\Delta_i$ through the sample.

Spatial heterogeneities already appear below the glass
transition in the equilibrium state at $\Phi=\Phi_0$, $\T=30\dg$C. This suggests that the suspension is already in a
'supercooled' regime, where a qualitative change in the character of
particles motion occurs. The crowding of the suspension is high enough
to induce small heterogeneity of dynamics in the sample, but not
enough to allow aging, out-of-equilibrium fall, or jamming to
occur\cite{AnnRevGlass-Ediger-2000}.
 
\begin{figure}
\includegraphics[width = 6.0cm]{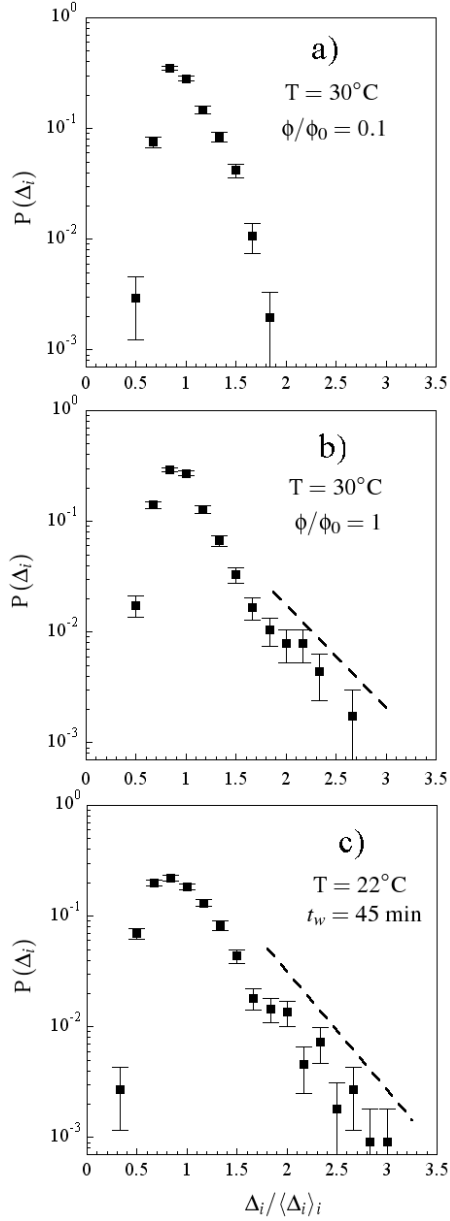}
\centering
\caption{Probability distribution functions of the variances
  $\Delta_i$, defined such as $\langle\Delta r^2_i
  (t)\rangle_{t'}=\Delta_i(t)$. (a) in the diluted liquid state at
  ($\T= 30\dg$C, $\Phi=0.1 \Phi_0$), over $1200$ probes, lag time $t=6
  \, 10^{-3}$ s; (b) in the reference liquid state at ($\T= 30\dg$C,
  $\Phi=\Phi_0$), over $1150$ probes, lag time $t=3.75 \,10^{-2}$ s;
  (c) in the glass state at $\T= 22 \dg$C and $t_w=45$ min, over
  $1220$ probes, lag time $t=6 \,10^{-2}$ s. (b,c): in the reference
  liquid state and in the glass state, the distribution is a Gaussian
  for short diffusion coefficients, and exhibits an exponential decay
  for large diffusion coefficients. No clear dependence of the
  distribution shape was found with the aging time. (a): in the
  diluted liquid state, the distribution is centered around a mean
  diffusion coefficient, and no exponential decay was observed.  }
\label{PDi}
\end{figure}

We now characterize the shape of the distributions $P(\Delta_i)$ of
the individual probes' mobilities, at equilibrium and
out-of-equilibrium. Fig. \ref{PDi} shows the spatial probability
distribution functions of the $\Delta_i$, defined such as
$\langle\Delta r^2_i (t)\rangle_{t'}=\Delta_i(t)$. In the reference
liquid state ($\Phi=\Phi_0$, $\T=30\dg$C) and in the glass state, the
distribution is a Gaussian for short diffusion coefficients, and
exhibits an exponential decay for large diffusion coefficients (Fig. \ref{PDi}(b,c)). In the
diluted liquid state ($\Phi=0.1 \Phi_0$, $\T=30\dg$C), the
distribution is centered around a mean diffusion coefficient, and no
exponential decay was observed for large diffusion coefficients (Fig. \ref{PDi}(a)). Given
a good statistics for the calculation of $\Delta_i(t)$, the shape and
behaviour of the distributions remains similar for various lag times $t$:
the distribution $P(\Delta_i)$ was found to develop an exponential
tail when approaching the glass state. In these experiments, no clear
dependence of the distribution shape was found with the aging time.

It is now straightforward to show that, given an exponential tail for
the distribution $P(\Delta_i)$, the resulting ensemble-averaged PDF
$p\left(\Delta x,t\right)$ will exhibit an exponential decay at large
displacements. We consider the ensemble-averaged PDF as the sum of
local Gaussian dynamics $p\left(\Delta x,t\right) = \int
\mathrm{d}\Delta_i\, P\left(\Delta_i\right) \exp{(-(\Delta x(t))^2 /
  \Delta_i)} / \sqrt{\pi \Delta_i}$, with a spatial distribution
$P\left(\Delta_i\right)=\frac{\exp\left(-\Delta_i/\Delta_0\right)}{\Delta_0}$,
at large $\Delta_i$. The larger the $\Delta_i$, the more it
contributes to the integral, so we can write~: $p\left(\Delta
x,t\right) \sim \int \mathrm{d}\Delta_i
\frac{\exp\left(-\Delta_i/\Delta_0\right)}{\Delta_0} \exp{(-(\Delta
  x(t))^2 / \Delta_i)} / \sqrt{\pi \Delta_i}$, which yields, using a
method of steepest descent, a leading term for the ensemble averaged
PDF at large displacements $p\left(\Delta x,t\right) \sim \exp\left( -2
|\Delta x(t)| / \sqrt{\Delta_0}\right)$. The exponential decay of the
distribution $P(\Delta_i)$, at large $\Delta_i$, results in an
exponential decay of the ensemble-averaged PDF.

The exponential decay of $P(\Delta_i)$ reveals that a small fraction
of tracers are intrinsically faster than average. Given that all the
individual PDF are Gaussian, the fast particles are responsible for
the large displacements in the ensemble-averaged PDF. Here, the
exponential decay of the ensemble-averaged PDF is thus the result of
the existence of faster mobility zones than average on our observation
time scales, rather than jumps experienced by the tracers from time to
time. Moreover, in these experiments, no evidence of temporally heterogeneous dynamics was
detected on the considered recording times, in the sense that fast
regions did not switch from rapid to slow dynamics over time (and {\it vice versa}).

We have characterized the statistical distribution of dynamical
environments, and show that it can explain the observed exponential
tail of the van Hove functions, in the concentrated regimes
(supercooled and glass). We analyze quantitatively below how the
variance $\sigma^2(t)$ of this distribution $P(\Delta_i)$ evolves when
approaching the glass transition, and beyond.

\subsubsection{Spatial heterogeneities when approaching the glass transition, and beyond}
\label{secvi}

A question that naturally arises
now to further characterize the spatial heterogeneities concerns the way
they evolve, when approaching and entering the glass
state. We now look at the
variance $\sigma^2(t)$ of the spatial distribution of the $\Delta_i(t)=\langle\Delta
r^2_i (t)\rangle_{t'}$ through the sample, given by :
\begin{equation*}
\frac{\sigma^2(\langle\Delta r^2_i (t)\rangle_{t'})}{\langle\Delta
  r^2 (t)\rangle^2_{t',i}}=\frac{\langle(\langle\Delta r^2_i
      (t)\rangle_{t'} - \langle\Delta r^2_i
      (t)\rangle_{t',i'})^2\rangle_i}{ \langle\Delta r^2
      (t)\rangle^2_{t',i} }
\end{equation*}
where $\langle\Delta r^2_i (t)\rangle_{t'}$ refers to the
temporal-averaged MSD for one probe, and $\langle\Delta r^2
(t)\rangle_{t',i}$ to the ensemble and temporal averaged MSD. This variance gives a measure of the intensity of spatial
heterogeneities in the sample. The larger the variance $\sigma^2$, the more distributed the variances $\Delta_i$ of the individual tracer motion through the sample, and the more the sample is spatially heterogenous.

\begin{figure}
\includegraphics[width = 8.0cm]{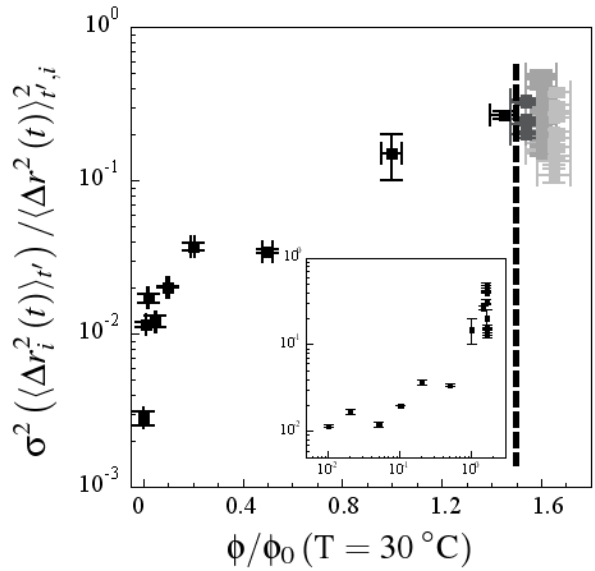}
\centering
\caption{Variance of the spatial distribution of the individual MSD as
  a function of the relative volume fraction, at equilibrium and in
  the glass states. The relative volume fraction is defined with
  respect to the volume fraction $\Phi_0$ of the reference equilibrium
  state at $\T=30\dg$C. Given a good statistics for the calculation of
  $\Delta_i(t)$, the variance $\sigma^2(t)$ was found to be
  independent of the lag time. The water represents the reference
  where no spatial heterogeneities are expected, and sets the
  sensitivity in our experiments. In thermosensitive suspensions, the
  value for $\sigma^2(t)$ is larger than the sensitivity, which
  indicates that spatial heterogeneities are present. At equilibrium,
  the variance increases with increasing volume fraction (roughly
  exponentially for large volume fractions, as a power law for low
  volume fractions, see inset), indicating the suspension becomes more
  and more spatially heterogeneous when approaching the glass
  transition. In the out-of equilibrium states (above the dashed
  vertical line), the suspension remains strongly spatially
  heterogeneous and also widely distributed (the different sets of
  points correspond to different quenches; within a quench, the
  different points correspond to different aging times). (inset):
  corresponding log-log plot.}
\label{vi}
\end{figure}

Figure \ref{vi} shows the variance $\sigma^2$ of the spatial
distribution of the $\langle\Delta r^2_i (t)\rangle_{t'}$ through the
sample, as a function of the volume fraction, before and after
entering the glass state. Starting from a very diluted liquid state, the
variance was found to increase with increasing volume fraction. Given
a good statistics for the calculation of $\Delta_i(t)$, the variance
$\sigma^2(t)$ was found to be independent of the lag time $t$, and Figure
\ref{vi} unchanged. The water represents the reference where no
spatial heterogeneities are expected, and sets the sensitivity in our
experiments. In thermosensitive suspensions, the value for
$\sigma^2(t)$ was found to be larger than the experimental
sensitivity, which indicates that spatial heterogeneities are present.
Figure \ref{vi} shows that the suspension dynamics becomes more and
more spatially heterogeneous when approaching the glass transition. As
seen in section \ref{sec-pdf}, the dynamics is spatially homogeneous
at low concentration ($\Phi/\Phi_0=0.1$, $\T=30\dg$C), and becomes
spatially heterogeneous, already below the glass transition
($\Phi/\Phi_0=1$, $\T=30\dg$C).

\begin{figure}
\includegraphics[width = 8.0cm]{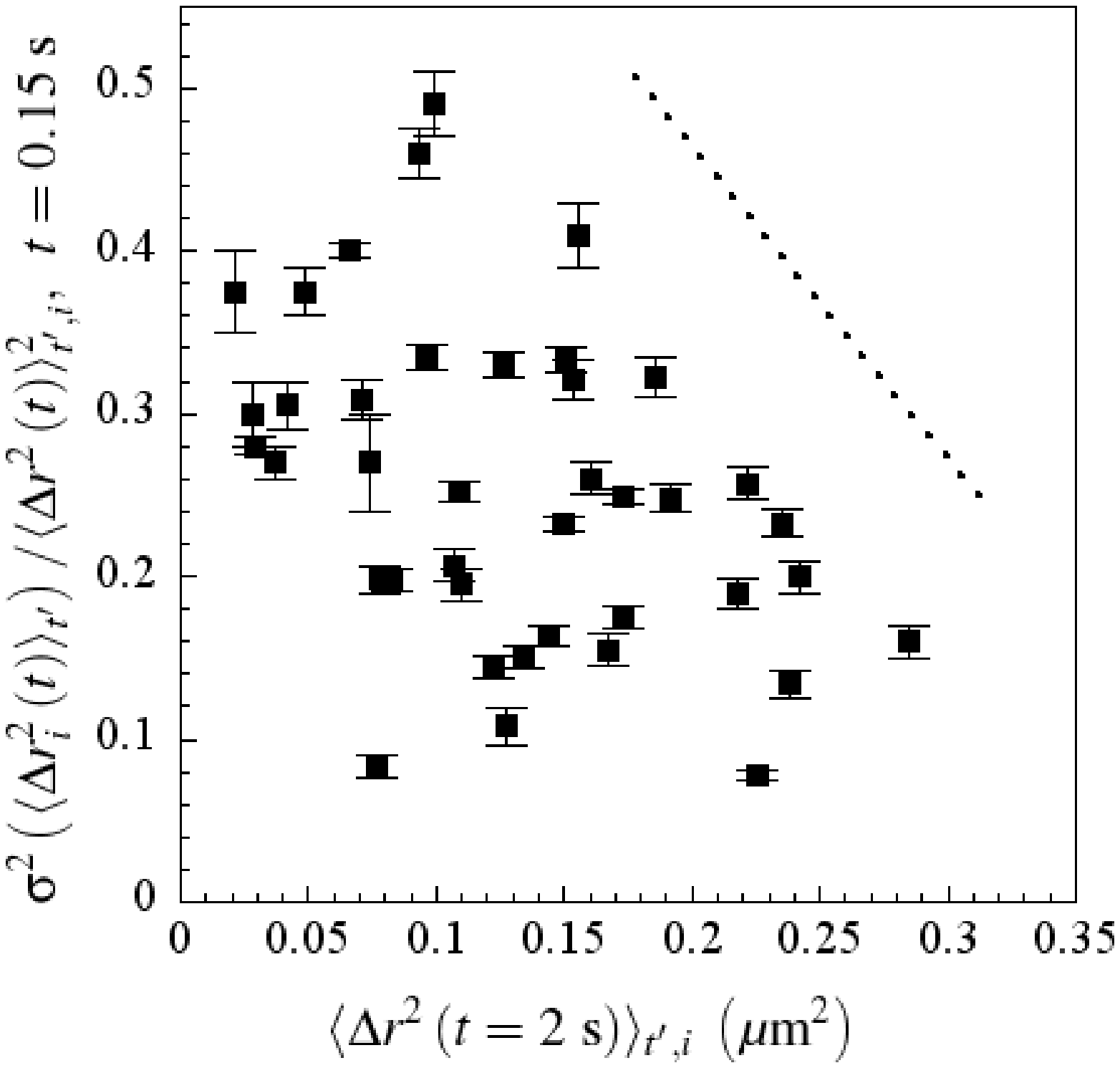}
\centering
\caption{Variance of the distribution of the individual MSD in the
  glass state, after a quench from $\T=30\dg$C to
  $\T=22\dg$C, as a function of a given MSD, at a lag time
  $t=2$~s. At $\T=30\dg$C, the suspension viscosity iss $6.0 \pm 0.3$~mPa.s. Although widely distributed, the variance $\sigma^2$ tends to decrease with the mean MSD, suggesting that
  the more the suspension is arrested the more the dynamics is
  spatially heterogeneous. The dashed line is a guide to the eye. }
\label{vi2}
\end{figure}

When entering the glass state (above the vertical line), the dynamics
remains strongly heterogeneous but also widely distributed. The
behaviour of spatial heterogeneities in the glass state during aging
is shown in Fig. \ref{vi2}, where we plot the variance $\sigma^2$ as a
function of a given MSD, $\langle\Delta r^2 (t)\rangle_{t',i}$, at
$t=2$~s. We take this MSD as a measure of how deep the suspension is
in the glass state after a quench, since this allows us to compare
quenches with different final temperatures and different realizations
of these quenches, from the same initial equilibrium state. For a
given $t$ in the diffusive regime or in the plateau, the variance
tends to increase with decreasing amplitude of the $\langle\Delta r^2
(t)\rangle_{t',i}$. This suggests that the deeper the suspension in an
arrested state, the more spatially heterogeneous the dynamics.

This work establishes the link
between the non Gaussian, exponentially tailed van Hove functions and
a spatial distribution of homogeneous dynamic environments in a soft
colloidal glass. The {\em intensity} of the dynamic heterogeneities is
characterized when approaching the glass transition. In aging experiments,
reproducibility in the out-of-equilibrium dynamics after a change in
the external control parameter, is a major experimental issue. In our
experiments, we noticed that taking the aging time $t_w$ as a
parameter of the 'glass state' gives qualitatively contradictory
results for $\sigma^2 (t_w)$, from one realization to the
other. However, choosing a given MSD, $\langle\Delta r^2
(t_0)\rangle_{t',i}$, as a measure of the 'glass state' leads to the
same qualitative behaviours for $\sigma^2$ as the suspension goes
deeper in the glass state. The dispersion of the data in
Fig. \ref{vi2} certainly comes from the poor statistics consecutive to
the use of tracers instead of the microgels themselves. More
systematic experiments in the aging regime, improving the statistics,
in suspensions with particles of various softness, are currently under
investigation.

Dynamic heterogeneities in the aging regime have been
investigated in a few
studies\cite{JPCM-Weeks-2003,PRE-Lynch-2008,PRL-Yunker-2009}, although
results in this regime might help to discriminate between
theories\cite{Berthier-review-2010}. Since the
particles are almost index matched and impossible to visualize
unambiguously on the microscope, the characterization of the shape and
size of the regions of homogeneous dynamic
environments were beyond the scope of this study,
as well as an attempt to question a link between the structure and the
dynamic heterogeneities\cite{PRL-Yunker-2009}. Here, the
dynamic heterogeneity is seen as a spatial heterogeneity on the
recording time scale, which was chosen to ensure quasi static
conditions. Complementary work will investigate the temporal
fluctuations of the dynamics and the relation between aging and temporally heterogeneous
dynamics.

\section{Conclusion}

We have investigated the aging dynamics and dynamic heterogeneity of a
soft thermosensitive microgels suspension, by looking at the thermal
fluctuations of tracers. By varying temperature, the volume fraction
could be tuned, and the characteristics of the equilibrium liquid
suspension and the out-of-equilibrium glass state during aging,
investigated {\em on the same sample}. 
Aging was observed over several hours by quenching the suspension from
an initial liquid state at high temperature. With increasing the
quench amplitude, the aging dynamics was found to be faster, and the system to enter
a deeper glass state.
 Spatially heterogeneous dynamics was observed to develop below the
glass transition, and to persist in the glass state. Individually, the
probes always exhibit a Gaussian PDF, whether at equilibrium or in the
glass state, revealing a locally homogeneous dynamical environment. With
increasing volume fraction, the local dynamical environment was found to change
significantly from one probe to another, sign of a spatially
heterogeneous dynamics.
In the supercooled regime,
characterized by a non aging heterogeneous system, and in the glass states, the statistical
distribution of these dynamical environments was shown to decay
exponentially, indicating that the dynamics is dominated by a small
fraction of faster mobility zones than average, rather than jumps
experienced by the tracers from time to
time, on the experimental timescales. We show that these distributions with exponential tails can explain the observed
exponential tails of the van Hove functions. 
Finally, we analyze the variance of these distributions, as a function
of the volume fraction, and as a function of the
age in the glass state. It gives a measure of the spatial
heterogeneity intensity in our samples. With increasing volume
fraction, spatially heterogeneous dynamics was found to amplify. In
the glass state, it tends to increase as the suspension gets more
arrested.

\section{Appendix}

\subsection{Cumulative Distribution Function}

The Cumulative Distribution Function (CDF) $\mathrm{C}\left(x\right)$
of a random variable ({\em e.g.} the displacement of a probe particle
during a fixed lag time, $\Delta x$) is the probability that this
random variable takes a value smaller or equal to $x$~:
$\mathrm{C}\left(x\right) = p\left(\Delta x \leq x \right)$. The CDF
of a random variable is deduced from its PDF $p$ by integration~:
$\mathrm{C}\left(x\right) = \int_{-\infty}^{x} \mathrm{d}x'
p\left(x'\right)$. A schematic representation of a typical CDF is
presented below~: {\begin{center}
    \includegraphics[width=4cm]{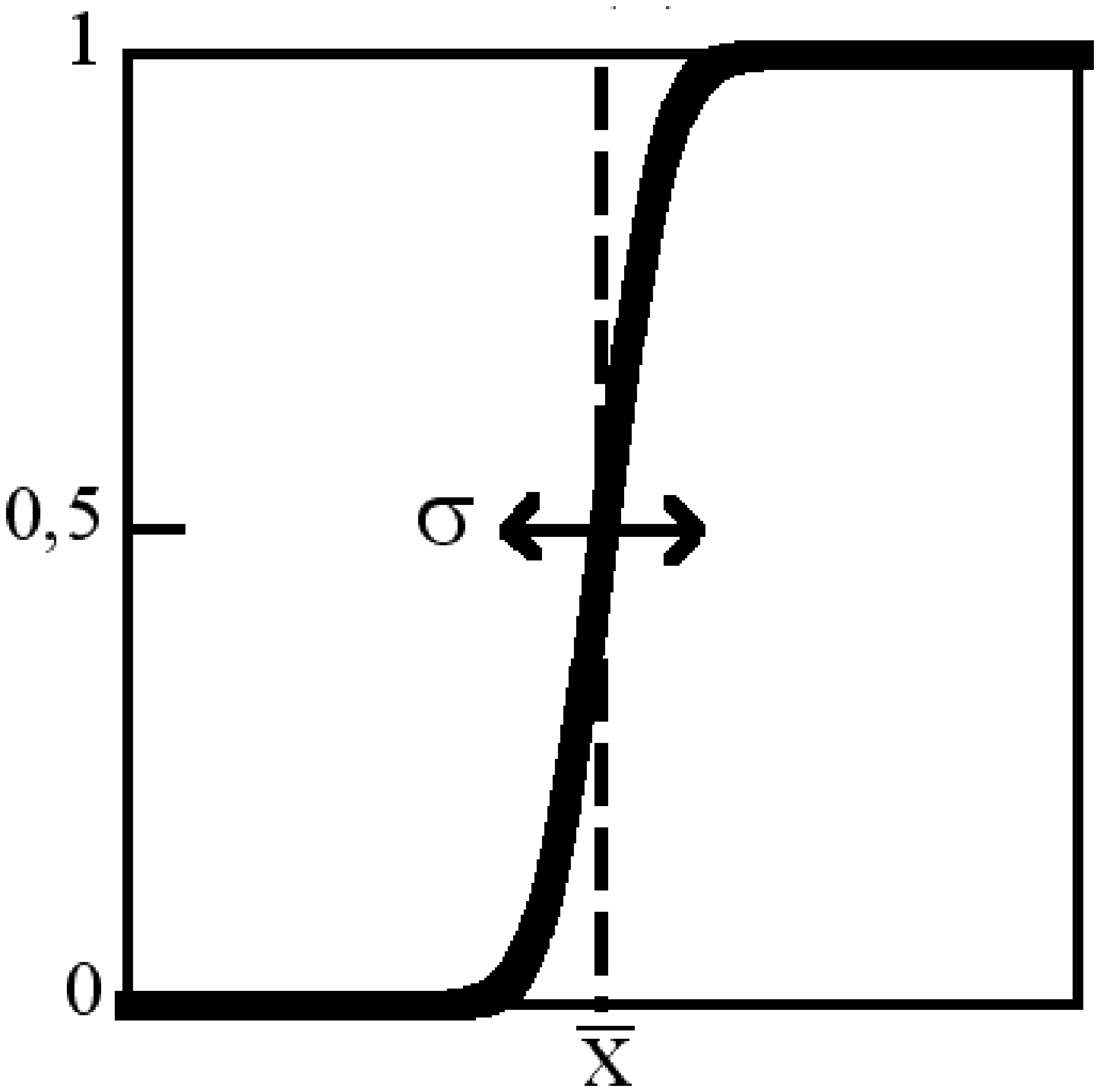}
\end{center}}
It monotonically increases from $0$ to $1$, is antisymmetric around
the variable median ($0$ in our case) and has a typical width of the
order of its standard deviation (square root of the MSD in our case).

Our interest here is to know the detailed shape of the {\em tails} of
the CDF, for large positive and negative displacements. Since the
situation is symmetrical, we focused on positive displacements
only. In order to zoom on what happens as $\mathrm{C}\left(x\right)$
reaches $1$, the tail is stretched by computing, for $x > 0$, the
quantity~:
\begin{equation}
y = -\ln\left(2\left(1-\mathrm{C}\left(x\right)\right)\right),
\label{ydef}
\end{equation}
which diverges as $\mathrm{C}\left(x\right)$ goes to $1$ ($x$ to
$+\infty$). For the negative tail, we computed $y' =
\ln\left(2\mathrm{C}\left(x\right)\right)$ and found results in
quantitative agreement with those of the positive tail.

Two particular cases were of great interest to us~: the case of a
Gaussian PDF of the displacements and the case of a PDF of
displacements decreasing exponentially for large displacements.\\ In
the case of a Gaussian pdf of the displacements $p\left(\Delta
x\right) = \exp\left(-(\Delta x)^2 / \langle\Delta r^2\rangle\right) /
\sqrt{\langle\Delta r^2\rangle}$, the CDF is given by~:
\begin{equation}
\mathrm{C}_{gauss}\left(x\right) = \frac{1+ \mathrm{erf}\left(x / \sqrt{\langle\Delta r^2\rangle}\right)}{2}.
\label{Cgauss}
\end{equation}
For large displacements ($x >> \sqrt{\langle\Delta r^2\rangle}$), the
stretched CDF is given by inserting Eq. \ref{Cgauss} in Eq.\ref{ydef}
and developping for large $x$ and yields~:
\begin{equation}
y_{gauss} = u^2 + \frac{1}{2} \ln \pi + \ln u + O\left(u^{-2}\right),
\end{equation}
where $u = x / \sqrt{\langle\Delta r^2\rangle}$. Thus a {\em
  quadratic} variation of $y$ with $x$ at leading order is reminiscent
of a {\em Gaussian} shape of the PDF of displacements. We fitted
variable $y$ in the Gaussian cases with its full expression
(Eq. \ref{Cgauss} inserted in Eq.\ref{ydef}) rather than the large
displacement development.

In the case of a PDF with exponential tails, if the PDF of
displacements follows the equation $p\left(\Delta x\right) =
\exp\left(-|\Delta x |/ \lambda\right) / \lambda$ for large $\Delta x$,
then The CDF of displacements is given for large $x$ by~:
\begin{equation}
\mathrm{C}_{exp} \left(x\right)= 1 - B e^{-\frac{x}{\lambda}},
\end{equation}  
where $B$ depends on the detailed shape of the PDF at small displacements. The stretched CDF is then given for large displacements by~:
\begin{equation}
y_{exp} = \frac{x}{\lambda} - \ln\left(2B\right).
\end{equation}
A {\em linear} divergence of the stretched CDF is thus equivalent to
an {\em exponential} decay of the PDF of displacements.

Finally, if the stretched CDF is fitted by $y = a \, u + b$ ($u =
x/\sqrt{\langle\Delta r^2\rangle}$), it is easily shown that the PDF
of displacements is given by
$\frac{a}{2}\exp\left(-b\right)\exp\left(-a\, u\right)$. If the
fit is valid on a range $u \in \left[u_0; u_1\right]$, then the PDF
decrease exponentially on a number $n$ of decades given by $n =
\log_{10} \left(p(u_0)/p(u_1)\right) = \left(a\left(u_1 -
u_0\right)\right) \log_{10}\left(e\right)$.

\begin{figure}
\includegraphics[width =
  8.5cm]{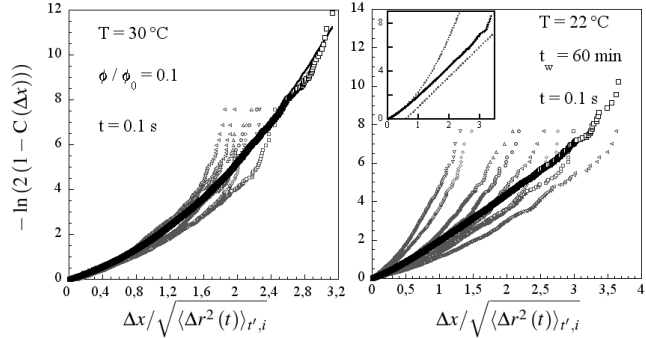} 
\centering
\caption{Cumulative distribution functions (CDF) of the probes'
  displacements in the liquid equilibrium state (left) and in the
  glass state (right). The grey symbols represent the individual CDF
  of the probes while the black symbols are for the ensemble-averaged
  CDF over the probes. In the liquid state, all the CDF are well
  described by a Gaussian distribution of displacements and the
  individual CDF are narrow distributed. In the glassy state, the
  spreading of individual Gaussian CDF is wide. The inset shows the
  ensemble-averaged CDF with two fits: a parabolic one at short
  displacements, and a linear one at large displacements, giving
  evidence that the displacements distribution is Gaussian-like for
  small displacement (as expected at first order) and exponential for
  large displacements.}
\label{CDF}
\end{figure}

Figure \ref{CDF} shows the individual and ensemble-averaged Cumulative
Distribution Functions (CDF) of the probes' displacements in the low
concentration liquid equilibrium state and in the glass state. In the
low concentration liquid state, the individual CDF are narrow
distributed, indicating a spatially homogeneous suspension. All the
CDF in the liquid state, individual and ensemble-averaged, could be
adjusted by a parabolic curve indicating a Gaussian behaviour of the
corresponding PDF. In the glass state, the individual
CDF are widely distributed and still described by a parabolic
function, as seen in Figure \ref{CDF}-right. Their ensemble-average
results in a linear function, giving evidence of an exponential decay
of the associated PDF over 3 decades. 

In the glass phase at $\mathrm{T}= 22^\circ\mathrm{C}$, we found $a =
2.40 \pm 0.05$ and $b = -0.40 \pm 0.05$, valid over a range $u \in
\left[0.7; 3.5\right]$. This is equivalent to an exponential decrease
of the PDF over $3$ decades.

\subsubsection*{Acknowledgements}
R. C. and B. A. are grateful to F. Van Wijland and C. Gay for very fruitful discussions. 

\bibliographystyle{unsrt} 
\bibliography{biblio-220111}

\begin{thebibliography}{10}

\bibitem{nature-Pusey-1986}
P.~N. Pusey and W.~van Megen.
\newblock Phase behavior of concentrated suspensions of nearly hard colloidal
  spheres.
\newblock {\em Nature}, 320:340--342, March 1986.

\bibitem{Science-vanBlaaderen-1995}
Alfons van Blaaderen and Pierre Wiltzius.
\newblock Real-space structure of colloidal hard-sphere glasses.
\newblock {\em Science}, 270:1177, November 1995.

\bibitem{Nat-Liu-1998}
Andrea~J. Liu and Sidney~R. Nagel.
\newblock Jamming is not just cool anymore.
\newblock {\em Nature}, 396:21, 1998.

\bibitem{Cipelletti-Weeks}
L.~Cipelletti and E.~Weeks.
\newblock Glassy dynamics and dynamical heterogeneities in colloids.
\newblock {\em arXiv:1009.6089v1[cond-mat.soft]}, 2010.

\bibitem{Mattson-Nature-2009}
J.~Mattson, H.~M. Wyss, A.~Fernandez-Nieves, K.~Miyazaki, Z.~Hu, D.~R.
  Reichman, and D.~Weitz.
\newblock Soft colloids make strong glasses.
\newblock {\em Nature}, 462:83, 2009.

\bibitem{ACIS-Saunders-1999}
Brian~R. Saunders and Brian Vincent.
\newblock Microgel particules as model colloids : theory, properties and
  applications.
\newblock {\em Advances in Colloid and Interface Science}, 80:1--25, 1999.

\bibitem{CPS-Senff-2000}
H.~Senff and W.~Richtering.
\newblock Influence of cross-link density on rheological properties of
  temperature-sensitive microgel suspensions.
\newblock {\em Colloid Polym. Sci.}, 278:830--840, February 2000.

\bibitem{Langmuir-Stieger-2004}
Markus Stieger, Jan~Skov Pederson, Peter Lindner, and Walter Richtering.
\newblock Are thermoresponsive microgels model systems for concentrated
  colloidal suspensions ? a rheology and small-angle neutron scattering study.
\newblock {\em Langmuir}, 20:7283--7292, June 2004.

\bibitem{CPS-Hellweg-2000}
T.~Hellweg, C.~D. Dewhurst, E.~Br\"uckner, K.~Kratz, and W.~Eimer.
\newblock Colloidal crystals made of poly(n-isopropylacrylamide) microgel
  particles.
\newblock {\em Colloid Polym. Sci.}, 278:972--978, March 2000.

\bibitem{JPCB-Lyon-2004}
L.~Andrew Lyon, Justin~D. Debord, Saet Byul~Debord, Clinton~D. Jones,
  Jonathan~G. McGrath, and Michael~J. Serpe.
\newblock Microgel colloidal crystals.
\newblock {\em Journal of Physical Chemistry B}, 108:19099--19108, May 2004.

\bibitem{Science-Alsayed-2005}
A.~M. Alsayed, M.F. Islam, J.~Zhang, P.J. Collings, and A.G. Yodh.
\newblock Premelting at defects within bulk colloidal crystals.
\newblock {\em Science}, 309:1207, August 2005.

\bibitem{PRE-Han-2008}
Y.~Han, N.~Y. Ha, A.~M. Alsayed, and A.~G. Yodh.
\newblock Melting of two-dimensional tunable-diameter colloidal crystals.
\newblock {\em Phys. Rev. E}, 77:041406, April 2008.

\bibitem{PRE-Lohr-2010}
M.~A. Lohr, A.~M. Alsayed, B.~G. Chen, Z.~Zhang, R.~D. Kamien, and A.~G. Yodh.
\newblock Helical packings and phase transformations of soft spheres in
  cylinders.
\newblock {\em Phys. Rev. E}, 81:040401, April 2010.

\bibitem{PRL-Purnomo-2008}
Eko~H. Purnomo, Dirk van~den Ende, Siva~A. Vanapalli, and Frieder Mugele.
\newblock Glass transition and aging in dense suspensions of thermosensitive
  microgel particles.
\newblock {\em Phys. Rev. Lett.}, 101(23):238301, Dec 2008.

\bibitem{Nature-Zhang-2009}
Z.~Zhang, N.~Xu, D.~T.~N. Chen, P.~Yunker, A.~M. Alsayed, K.~B. Aptowicz,
  P.~Habdas, A.~J. Liu, S.~R. Nagel, and A.~G. Yodh.
\newblock Thermal vestige of the zero-temperature jamming transition.
\newblock {\em Nature}, 459:07998, May 2009.

\bibitem{PRE-Purmono-2010}
Dirk van~den Ende, Eko~H. Purmono, Michel H.~G. Duits, Walter Richtering, and
  Frieder Mugele.
\newblock Aging in dense suspensions of soft thermoresponsive microgel
  particles studied with particle-tracking microrheology.
\newblock {\em Phys. Rev. E}, 81:011404, January 2010.

\bibitem{Langmuir-Kasper-1998}
A.~Kasper, E.~Bartsch, and H.~Sillescu.
\newblock Self-diffusion in concentrated colloid suspensions studied by digital
  video microscopy of core-shell tracer particles.
\newblock {\em Langmuir}, page 5004, 1998.

\bibitem{PRE-Marcus-1999}
A.~H. Marcus, J.~Schofield, and S.A. Rice.
\newblock Experimental observations of non-gaussian behavior and stringlike
  cooperative dynamics in concentrated quasi-two-dimensional collidal liquids.
\newblock {\em Phys. Rev. E}, 60:5725, 1999.

\bibitem{Science-Kegel-2000}
Willem~K. Kegel and Alfons van Blaaderen.
\newblock Direct observation of dynamical heterogeneities in colloidal
  hard-sphere suspensions.
\newblock {\em Science}, 287:290, January 2000.

\bibitem{Science-Weeks-2000}
Eirc~R. Weeks, John~C. Crocker, Andrew~C. Levitt, Andrew Schofield, and
  David~A. Weitz.
\newblock Three-dimentional direct imaging of structural relaxation near the
  colloidal glass transition.
\newblock {\em Science}, 287:627, January 2000.

\bibitem{PRL-Cloitre-2000}
M.~Cloitre, R.~Borrega, and L.~Leibler.
\newblock Rheological aging and rejuvenation in microgel pastes.
\newblock {\em Phys. Rev. Lett.}, 85:4819, 2000.

\bibitem{PRE-Lynch-2008}
J.M. Lynch, G.C. Cianci, and E.R. Weeks.
\newblock Dynamics and structure of an aging binary colloidal glass.
\newblock {\em Phys. Rev. E}, 78:031410, 2008.

\bibitem{PRL-Yunker-2009}
P.~Yunker, Z.~Zhang, K.~B. Aptowicz, and A.~G. Yodh.
\newblock {\em Phys. Rev. Lett.}, 103:115701, 2009.

\bibitem{PRB-Zorn-1997}
R.~Zorn.
\newblock Deviation from gaussian behavior in the self-correlation function of
  the proton motion in polybutadiene.
\newblock {\em Phys. Rev. B}, 55:6249, 1997.

\bibitem{AnnRevGlass-Ediger-2000}
M.D. Ediger.
\newblock Spatial heterogeneous dynamics in supercooled liquids.
\newblock {\em Annual Review of Physical Chemistry}, 51:99--128, 2000.

\bibitem{Nature-Vidal-2000}
E.~Vidal~Russel and N.~E. Israeloff.
\newblock Direct observation of molecular cooperativity near the glass
  transition.
\newblock {\em Nature}, 2000.

\bibitem{PRL-Marty-2005}
G.~Marty and O.~Dauchot.
\newblock Subdiffusion and cage effect in a sheared granular material.
\newblock {\em Phys. Rev. Lett.}, 94(1):015701, Jan 2005.

\bibitem{PRE-Abate-2007}
A.~R. Abate and D.~J. Durian.
\newblock Topological persistence and dynamical heterogeneities near jamming.
\newblock {\em Phys. Rev. E}, 76:021306, 2007.

\bibitem{PRL-Doliwa-1998}
B.~Doliwa and A.~Heuer.
\newblock Cage effect, local anisotropies, and dynamical heterogeneities at the
  glass transition: a computer study of hard spheres.
\newblock {\em Phys. Rev. Lett.}, 80:4915, 1998.

\bibitem{JNCS-Glotzer-2000}
S.~C. Glotzer.
\newblock Spatially heterogeneous dynamics in liquids: insights from
  simulation.
\newblock {\em Journ. of Non-crystalline Solids}, 274:342, 2000.

\bibitem{PRE-Elmasri-2010}
D.~El~Masri, L.~Berthier, and L.~Cipelletti.
\newblock Subdiffusion and intermittent dynamic fluctuations in the aging
  regime of concentrated hard spheres.
\newblock {\em Phys. Rev. E}, 82:031503, 2010.

\bibitem{JPCM-Richert-2002}
Ranko Richert.
\newblock Heterogeneous dynamics in liquids : fluctuations in space and time.
\newblock {\em Journal of Physics : Condensed Matter}, 14:R703--R738, May 2002.

\bibitem{PRL-Chaudhuri-2007}
Pinaki Chaudhuri, Walter Kob, and Ludovic Berthier.
\newblock Universal nature of particle displacements close to glass and jamming
  transitions.
\newblock {\em Phys. Rev. Lett.}, 99:060604, august 2007.

\bibitem{JNCS-Richert-1994}
R.~Richert.
\newblock {\em Journ. of Non-crystalline Solids}, 172:209, 1994.

\bibitem{PRE-Vorselaars-2007}
B.~Vorselaars, A.V. Lyulin, K.~Karatasos, and M.A.J. Michels.
\newblock Non-gaussian nature of glassy dynammics by cage to cage motion.
\newblock {\em Phys. Rev. E}, 75:011504, 2007.

\bibitem{Sessoms-2009}
D.A. Sessoms, I.~Bischofberger, L.~Cipelletti, and V.~Trappe.
\newblock {\em Phil. Trans. R. Soc. A}, 367:5013 --5032, 2009.

\bibitem{Berne_Pecora}
Bruce~J. Berne and Robert Pecora.
\newblock {\em Dynamic light Scattering With Application to Chemistry, Biology
  and Physics}.
\newblock Dover Publications, Inc., 2000.

\bibitem{JCP-Tanaka-1979}
T.~Tanaka and D.~J. Filmore.
\newblock Kinetics of swelling of gels.
\newblock {\em Journ. of Chem. Phys.}, 70:1214--1218, 1979.

\bibitem{epl-Lechenault-2008-1}
Fr\'ed\'eric Lechenault, Olivier Dauchot, Giulio Biroli, and J.P. Bouchaud.
\newblock Critical scaling and heterogeneous superdiﬀusion across the
  jamming/rigidity transition of a granular glass.
\newblock {\em Europhys. Lett.}, 83(4):46003, 2008.

\bibitem{PRE-Abou-2001}
B\'ereng\`ere Abou, Daniel Bonn, and Jacques Meunier.
\newblock Aging dynamics in a colloidal glass.
\newblock {\em Phys. Rev. E}, 64(2):021510, July 2001.

\bibitem{PRE-Bellour-2003}
Knaebel~A. Bellour, M.~and, J.~L. Harden, F.~Lequeux, and J.-P. Munch.
\newblock Aging processes and scale dependence in soft glassy colloidal
  suspensions.
\newblock {\em Phys. Rev. E}, 67(3):031405, March 2003.

\bibitem{PRE-Kaloun-2005}
S.~Kaloun, R.~Skouri, M.~Skouri, J.~P. Munch, and F.~Schosseler.
\newblock Successive exponential and full aging regimes evidenced by tracer
  diffusion in a colloidal glass.
\newblock {\em Phys. Rev. E}, 72(1):011403, July 2005.

\bibitem{RPP-Waigh-2005}
T.~A. Waigh.
\newblock Microrheology of complex fluids.
\newblock {\em Rep. Prog. Phys.}, 68:685--742, 2005.

\bibitem{JPCM-Weeks-2003}
Rachel~E. Courtland and Eric~R. Weeks.
\newblock Direct visualization of ageing in colloidal glasses.
\newblock {\em Journal of Physics : Condensed Matter}, 15:S359--S365, January
  2003.

\bibitem{Berthier-review-2010}
L.~Berthier, G.~Biroli, J.-P. Bouchaud, and R.~L. Jack.
\newblock Overview of different characterisations of dynamic heterogeneity.
\newblock {\em arXiv:1009.4765v2[cond-mat.soft]}, 2010.

\end{thebibliography}

\end{document}